\documentclass[aps,amsmath,amssymb]{revtex4}
\usepackage{graphicx}
\usepackage{hyperref}
\usepackage{graphicx}

\usepackage[utf8]{inputenc}
\usepackage[T1]{fontenc}
\usepackage{tabularx}

\newenvironment{conditions*}
  {\par\vspace{\abovedisplayskip}\noindent
   \tabularx{\columnwidth}{>{$}l<{$} @{${}={}$} >{\raggedright\arraybackslash}X}}
  {\endtabularx\par\vspace{\belowdisplayskip}}

\def\si{\sigma}

\def\pa{\partial}

\def\11{{\mathbb 1}}

\def\re{{\rm e}}

\def\beq{\begin{equation}}
\def\eeq{\end{equation}}
\def\bea{\begin{eqnarray}}
\def\eea{\end{eqnarray}}
\def\nn{\nonumber}

\def\Ra{\Rightarrow}

\begin{document}

\title{Ising model with a magnetic field}
\author{Krzysztof A. Meissner$^1$,\\ Dobiesław Ircha$^2$, Wojciech Olszewski$^2$, Joanna Ruta$^2$,
Anna Słapek$^2$}
\affiliation{$^1$Faculty of Physics,
University of Warsaw\\
Pasteura 5, 02-093 Warsaw, Poland\\
$^2$Scott Tiger SA, Kolektorska 10, 01-692 Warsaw, Poland}

\begin{abstract} 
\noindent
The paper presents the low temperature expansion of the 2D Ising model in the presence of the magnetic field  in powers of $x=\exp(-J/(kT))$ and $z=\exp(B/(kT))$ with full polynomials in $z$ up to $x^{88}$ and full polynomials in $x^4$ up to $z^{-60}$, in the latter case the polynomials are explicitly given. The new result presented in the paper is an expansion not in inverse powers of $z$ but in $(z^2+x^8)^{-k}$ where the subsequent coefficients (polynomials in $x^4$) turn out to be divisible by increasing powers of $(1-x^4)$. The paper describes both the analytic expansions of the partition function and the efficient combinatorial methods to get the coefficients of the expansion.
\end{abstract}
\maketitle

\section{Introduction}
\noindent
The Ising model was proposed by Lenz in 1920 and solved in one dimension, therefore without the phase transition, by E. Ising in 1925 in his doctoral dissertation \cite{EI}. The main result is due to L. Onsager \cite{LO} in 1944 where the exact partition function for the model in 2D in the absence of the magnetic field $B$ was calculated. In 1952 C.N. Yang has proven an exact formula (earlier announced by L. Onsager and B. Kaufman in 1949)  for the first derivative of the partition function of the Ising Model in 2D with respect to $B$ at $B=0$ (magnetization). There are thousands of papers on the subject trying to include the non-vanishing magnetic field and huge body of results, both numerical and analytical, exists for the Ising on finite lattices. The paper presents the largest (as far as we know) full low temperature expansion in the presence of the magnetic field in powers of $x=\exp(-J/(kT))$ and $z=\exp(B/(kT))$ describing both the analytic expansions of the partition function and the efficient combinatorial methods to get the coefficients of the expansion in $x^4$ (as polynomials in $z^{-2}$ and in $z^{-2}$ (as polynomials in $x^4$) in the former case up to $x^{88}$ and in the latter up to $z^{-60}$. In the combinatorial part new algorithms were invented to speed up the computations and the supercomputing power of the Grid.pl was used with parallel computing using 64 GPU cards (in the GPU oriented version of the algorithm) or 15000 CPU (in the CPU oriented version).

\section{Basic considerations}
\noindent
On a 2D square lattice with $N$ 'spins' $\si= \pm 1$ we introduce a hamiltonian
\beq
H = -\frac{J}{2} \sum_{i,j,i\ne j}\si_i\si_j -B\sum_i\si_i.
\eeq
where the first sum runs over closest neighbours only. We will assume that the system is on a square $\sim \sqrt{N}\times \sqrt{N}$ with periodic boundary conditions. We will introduce the notation
\beq
x:=\re^{-\beta J},\ \ \ \ z:=\re^{\beta B}
\eeq
and assume that $x\leqslant 1$ and $z\geqslant 1$.
 
We define the free energy
\beq
\re^{-\beta NF_N} = Z_N =\sum\re^{-\beta (H-H_0)}
\eeq
where the sum runs over all configurations and $H_0$ is the lowest energy corresponding to the configuration with all spins pointing in the direction of $B$. The goal is to calculate the free energy per spin $F_N$ in the thermodynamical
limit $N\to \infty$. 

A simple case of $J = 0$ gives immediately the result
\beq
-\beta F_N(J = 0) = \ln\left(1 +\frac{1}{z^2}\right)
\eeq
The low temperature expansion starts from a special configuration with all spins directed in the direction of $B$ and the subsequent terms in the expansion are derived from configurations with more and more
inverted spins. For $N$ sufficiently large so that the periodic boundary conditions do no not play a role at a given order we have
\bea
\re^{-\beta NF_N} &=& \left(1 + \frac{N}{z^2}x^8 + \frac{N}{2z^4}\left((N-5)x^{16}+4x^{12}\right)+
 \frac{N}{6z^6}\left((N^2-15N+62)x^{24}+12(N-6)x^{20}+36x^{16}\right)+\ldots\right) \nn\\
&=&\left(1 + \frac{x^8}{z^2} + \frac{x^{12}(1-x^4)2}{z^4}+\frac{x^{16}(1-x^4)(-8x^{4}+6)}{z^6}+\ldots\right)^N
\label{partf}
\eea
where the polynomial in brackets is finite with the last term equal to $z^{-2N}$.

The task is to calculate the thermodynamical limit of the expression in brackets
\beq
\exp(f^\infty(x, z)) =\lim_{N\to\infty}\exp(-\beta F_N(x, z)) =\exp\left(\sum_{m,k}C_{m,k}\,x^{4m}z^{-2k}\right)
\eeq
For example
\beq
\exp(f^\infty(1,z)) = 1 +\frac{1}{z^2}
\eeq
The famous result of Onsager \cite{LO} gives the full result for the case $z=1$ ($B = 0$):
\beq\label{Onsresult}
\exp(f^\infty_{Ons}(x)) = \lim_{z\to1}\exp(f^\infty(x, z)) = (1 + x^4)\exp\left( -
\sum_{n=1}^{\infty}\left(\frac{(2n)!}{(n!)^2}\right)^2
\frac{1}{4n}\left(\frac{y}{4(1 + y)^2}\right)^n\right)
\eeq
where
\beq
y =\frac{4}{\left(\frac{1}{x^2}-x^2\right)^2}
\eeq
In the original article the result was expressed in elliptic functions. This result serves as a check for our results. The
beginning of the expansion:
\beq
\exp(f^\infty_{Ons}(x)) = 1 + x^8+2x^{12}+5x^{16} +14x^{20}+44x^{24}+152x^{28}+566x^{32}+\ldots
\label{Onsexplicit}
\eeq
The result of Yang \cite{CNY} for the magnetization at $B = 0$
\beq
\left. z\frac{\pa f^\infty(x,z)}{\pa z}\right|_{z=1}=(1-y^2)^{\frac18}-1
\label{Yang}
\eeq
although extremely simple was obtained by a very complicated method and it serves as yet another check on the
results. The result for susceptibility $\left.z\pa_z(z\pa_z (f^\infty(x,z)))\right|_{z=1}$ is not known analytically but only as a beginning of an expansion in $x^4$ (there is a large literature on the calculation of susceptibility, for example in \cite{IGE} one can find explicitly 38 first coefficients of the expansion).

The phase transition (for $B=0$) occurs for a temperature when the bracket in (\ref{Onsresult}) diverges:
\beq
y = 1 \Ra  x_c^4+x_c^{-4}=6\Ra x_c^4=3 - 2\sqrt{2}
\eeq
The description of phase transition when $B\ne 0$ is not analytically known.

\section{Combinatorial and symbolic algorithms}
\noindent
We consider a square lattice $L\times L$ on the torus and define configuration $\sigma$ 
as any assignment of values $\sigma_{i, j}$, spanning the set $\{ 0, 1 \}$, 
to the lattice nodes $i,j$ $\in$  [0, L-1].
Let $C(\sigma)$ be a circumference function on $\sigma$, defined as follows:

\beq 
   C(\sigma) = \sum_{i=0}^{L-1} \sum_{j=0}^{L-1} (\sigma_{i,j} \oplus \sigma_{i,j+1\!\!\!\!\! \mod L} + \sigma_{i,j} \oplus \sigma_{ i + 1 \!\!\!\!\! \mod L, j} )
\eeq
where $\oplus$ denotes exclusive OR. 
Finding the formula proposed in this work requires
determining the number of configurations $\sigma$ with a given value of $C(\sigma)$ 
for a fixed number $m$ of ones in the $\sigma_{i,j}$ set and a given torus size $L$.
The problem is computationally extensive, especially in the parameter range of interest. The total number of such configurations 
for a given $m$ and $L$ is $\binom{L^2}{m}$.
Below we present two variants of an algorithm for finding the required number of configurations and  describe their respective differences.  

\vspace{2mm}
\begin{center}  
    Algorithm variant $I$ 
\end{center}

We denote by $a$ the number of configurations $\sigma$ sharing the same value of $q=C(\sigma)$, 
for a given m and L. The total number of configurations for $m=10$ and $L=21$ is $69180774489220679208$. 
Sample values of $q$ and $a$ are shown in the table below.

\begin{center} 
    \begin{tabular}{ c r } 
         q & a \\
    \hline     
        14  &  13230 \\ 
        16  &  397782 \\ 
        18  &  17229870 \\ 
        $\cdots$ & $\cdots$ \\ 
        36  &  3728300274453675564 \\ 
        38  &  19491600033692972892 \\ 
        40  &  45521466242717189340 \\ 
    \end{tabular} 
\end{center}

The main idea of the proposed algorithm is to represent each configuration $\sigma$ as 
an $L$-element sequence $p = [ p_1, p_2, \ldots p_L ]$ of sums over the rows (or columns) of the torus 
lattice, i.e. $\sum_{i=1}^{L} p_i = m, p_i \in \{0, 1, \ldots L\}$. Thus, each element of the sequence 
is equal to the number of ones in the corresponding row (or column).
Considering that the studied values of $m$ and $L$ are of the same order, there are many 
sequences $p$ with multiple zero values.
We introduce a polynomial $g$:

\beq  
    g(x) = \sum_{i} a_i  x^{q_i}
\eeq    
where:
\begin{conditions*}
    q_i & the value of $C(\sigma)$ \\
    a_i & the number of configurations $\sigma$ with a given value of $q_i$
\end{conditions*}

We observe, that the number of configurations having the same value of the $C(\sigma)$ 
function can be calculated by determining the polynomial $g$.
Further, we find that the polynomial $g(x)$ can be formulated as an expression of polynomials $g_{k}(x)$ 
representing all subsequences $G_k$ of the sequence $p$ separated by one or more consecutive zeros or ones. 
For the separator 0, this expression is the product of polynomials representing individual subsequences $G_k$, 
while for the separator 1 an analogous algebraic formula can be given. 

For sequences with a size equal to the size of the torus $L$, the algorithm should be run for all the sequences 
of length $p_0$ and the computation of the final output should take into account the correspondence of the values 
for $p_0$ and $p_{L-1}$. Example: for  $ L = 21 $ and sequence $ p = [2, 3, 7, 5, 3] $, the polynomial representation is of the form:

\bea  
g(x) &=& 1890 x^{24}+35469 x^{26}+513429 x^{28}+6059214 x^{30}\nn\\
&&+60282390 x^{32}+532295715 x^{34}\nn\\
&&+4216243920 x^{36}+30319386972 x^{38}\nn\\
&&+198904938603 x^{40}+1191973363014 x^{42}\nn\\
&&+6515264362656 x^{44}+32349141844341 x^{46}\nn\\
&&+145002536779275 x^{48} + 581984976967422 x^{50}\nn\\
&&+2071038766883856 x^{52}+6459065687883018 x^{54}\nn\\
&&+17419061247080493 x^{56} +40019991455704323 x^{58}\nn\\
&&+77114090303006253 x^{60} +122702527182508161 x^{62}\nn\\
&&+158880959047544181 x^{64} +165175913830079061 x^{66}\nn\\
&&+136114567514444358 x^{68} +87688385553126582 x^{70}\nn\\
&&+43396295154271416 x^{72}+16078044280858812 x^{74}\nn\\
&&+4267980947851488 x^{76}+742151135328120 x^{78}+65807307035568 x^{80} 
\eea

Finding polynomial representations
$g_{k}(x)$ for all $G_k$ groups is computationally extensive. The naive approach requires reviewing all 
configurations that make up a $G_k $ group. Work is underway to develop a more efficient 
algebraic method reducing the search space of sequences and number of matrix operations. 

We can observe that the generating function $g(x)$ is invariant to certain transformations 
of the subsequences $G_k$. These transformations include:
\begin{itemize}
    \item \textbf{rotation} - In torus topology, a sequence can be started at any given index; All such sequences are equivalent,
     e.g. ($p_0$, $p_1$, ..., $p_j$, ... $p_m$) is equivalent to
     ($p_j$, ... $p_m$, $p_0$, ..., $p_{j-1}$)
     \item \textbf{inversion} of any subsequence $G_k$ - Any subsequence ($m_0$ ... $m_{j-1}$, $m_j$) is equivalent to subsequence ($m_j$, $m_{j-1}$ ... $m_0$)
     \item \textbf{translation} - Elements of the zero subsequences (separators) can be shifted between those subsequences, as long as each separator subsequence contains at least one element. 
     \item \textbf{swap} - All sequences constructed by swapping entire subsequences $G_k$ are equivalent to the original sequence $p$.
\end{itemize}

Given a sequence, we can formulate an algorithm for finding the number of such transformations. Moreover, we can designate a 
sequence representing the group of sequences sharing the same value of $C(\sigma)$,  resulting from invariant transformations, 
and use it in downstream computations. Intermediate results for subsequences $G_k$ shared across groups $p$ are reused to reduce computational complexity. 
The final result for a sequence $p$ is calculated based on the values found for all its subsequences $G_k$.

\vspace{2mm}
\begin{center}  
Variant II 
\end{center}

In this algorithm variant we fix the values of L and q=C( $\sigma$ ) and look for the number of configurations with $m$ ones, 
for all values of $m$. We note that for a given value of $C$, we can reduce computational complexity by performing the calculations 
on a lattice of size $\tfrac{C}{2}$ + $1$.

Below is an example for $C=36$ i $L=19$

\begin{center} 
    \begin{tabular}{ c r } 
		81  &  361 \\
		80  &  2166 \\
		79  &  7942 \\
        $\cdots$ & $\cdots$ \\ 
		11  &  586185684966484668 \\
		10  &  682914380225164860 \\
		9   &  172190699515632837 \\
    \end{tabular} 
\end{center} 

Next, for a perimeter $C$, we define $C_v$ and $C_h$, denoting, respectively, the number of vertical and horizontal lattice edges on $C$:

\begin{center} $C_h = [C_{h1}, C_{h2}, \ldots, C_{hn}] $
\end{center} 

where $C_{hi}$ is the number of horizontal edges in column ~$i$. \\

Similarly to variant I, we can use symmetries in sequences $C_v$ and $C_h$ to reduce the complexity of the algorithm 
(in this method for separator $0$ only). We observe that these symmetries allow to perform computations for~$C_h$ in the 
range of $1, \ldots,\lfloor{\frac{C}{4}}\rfloor$ and reuse the results for higher values. \\
The above algorithm can be implemented as separate tasks performed on a computing cluster. 
These tasks, i.e., finding symmetries, calculating the value of the polynomial generating function and 
construction of the final solution are easily parallelizable and can be run on GPUs.

\section{Results}
\noindent
We assume that $x \leq 1$ and $z > 1$ so all the expansions are in positive powers of $x$ and negative powers of $z$. We
will present the results in two different expansions: all powers of $x$ up to a given inverse power of $z$ (in our case up
to $z^{-60}$) and all inverse powers of $z$ up to a given power of $x$ (in our case up to $x^{88}$).

\subsection{Expansion in inverse powers of $z$}

As it turns out it is much better to expand in inverse powers of $z^2 +x^8$ and not $z^2$. It is a new result that the polynomials being coefficients of the expansion are divisible by the growing powers of $(1-x^4)$:
\beq\label{Mpol}
z^2 \exp(f^\infty) = z^2 + x^8 +\sum_{k=1}^\infty
\frac{(1 - x^4)^{[k/2]+1}x^{4l}M_k(x^4)}{(z^2 + x^8)^k}
\eeq
where $M_k(x^4)$ are ’magnetic’ polynomials of order $(2k + 1 - [k/2] - l)$ with integer coefficients and
\beq
l = [\sqrt{4k + 3}] + 1
\eeq
In the Appendix all $M_k$ polynomials are given up to $k=29$ what corresponds to $\exp(f^\infty)$ up to $z^{-60}$. Thanks to the powers of $x$ in front of $M_k$ in (\ref{Mpol}) the number of polynomial coefficients to be calculated at each order of $z^{-2k}$ is reduced by $([k/2]+1)$. 

The fact that the coefficients have to be divisible by the growing power of $(1-x^4)$ is a very powerful check on calculations. 

\subsection{Expansion in powers of $x$}
The expansion reads
\beq
\exp(f^\infty) = 1 +\sum_{k=1}^{\infty}\frac{x^{8k}}{z^{2k^2}} P_k^s(z^2) +
2\sum_{k=1}^{\infty}\frac{x^{8k+4}}{z^{2k(k+1)}} P_k^r(z^2) 
\eeq
The superscripts $s$ and $r$ stand for ’square’ and ’rectangle’ since the first class starts from a square $k \times k$ and the
subsequent powers of $z^2$ correspond to reversing spins keeping the same perimeter and the second class starts from two
rectangles $k\times(k+1)$ and $(k+1)\times k$ and again the subsequent powers of $z^2$ correspond to reversing spins keeping the same perimeter.

We quote first few polynomials  (they were calculated up to $P^s_{11}$ and $P^r_{10}$ but they are too long to be presented here)
\bea
P^s_1 &=& 1\\
P^s_2 &=& 1 + 6z^2 - 2z^4\nn\\
P^s_3 &=& 1 + 6z^2 + 22z^4 + 40z^6 + 44z^8 - 77z^{10} + 8z^{12}\nn\\
P^s_4 &=& 1 + 6z^2 + 22z^4 + 68z^6 + 151z^8 + 310z^{10} + 462z^{12}\nn\\
&&+546z^{14} + 221z^{16} - 424z^{18}- 1556z^{20} + 799z^{22} - 40z^{24}\nn\\
P^s_5&=&1 + 6z^2 + 22z^4 + 68z^6 + 187z^8 + 426z^{10} +914z^{12}+1728z^{14}+2979z^{16}\nn\\
&& +4572z^{18} +6426z^{20} +7444z^{22}+7557z^{24}+3699z^{26}-2696z^{28}-14444z^{30}\nn\\
&&-18964z^{32}-13598z^{34}+30348z^{36}-7672z^{38}+225z^{40}
\eea
The polynomials $P^s_k$ are of order $k(k - 1)$ in $z^2$.

The second set:
\bea
P^r_1 &=& 1\\
P^r_2 &=& 1 + 4z^2 + 9z^4 - 7z^6\nn\\
P^r_3 &=& 1 + 4z^2 + 15z^4 + 36z^6 + 67z^8 + 69z^{10} + 20z^{12} - 185z^{14} + 49z^{16}\nn\\
P^r_4 &=& 1 + 4z^2 + 15z^4 + 44z^6 + 109z^8 + 228z^{10} + 432z^{12} + 671z^{14} + 951z^{16}\nn\\
&&+948z^{18} + 615z^{20} - 611z^{22} - 1854z^{24} - 2735z^{26} + 2652z^{28} - 353z^{30}\nn\\
P^r_5&=&1 + 4z^2 + 15z^4 + 44z^6 + 119z^8 + 280z^{10} +604z^{12}+1204z^{14}+2236z^{16}+3787z^{18}\nn\\
&&  +6088z^{20} +8873z^{22}+12000z^{24}+14487z^{26}+15697z^{28}+12729z^{30}+6636z^{32}\nn\\
&&-8736z^{34}-24893z^{36}-43974z^{38}-32931z^{40}-347z^{42}+75787z^{44}-32552z^{46}+2602z^{48}
\eea
The polynomials $P^r_k$ are of order $(k^2 - 1)$ in $z^2$.

The lowest coefficients in these polynomials are straightforward to understand – for example
6 in $P^s_k$ corresponds to removing four corners of the square $k\times k$ plus two rectangles $(k - 1) \times (k + 1)$ 
and 4 in $P^r_k$ corresponds to removing four corners of the rectangle.

These polynomials for $z = 1$ should match the expansion (\ref{Onsexplicit}) and the derivatives $\pa_z$ at $z=1$ the expansion (\ref{Yang}) what are useful checks on the calculation.

We can note that some patterns emerge (what seems to be related to \cite{Baxter}). For $P^s$ we have
\beq
A^s =\prod_{m=1}^{\infty}(1 - z^{2m})^{-3}(1 + z^{2m})(1 + z^{4m-2})^2 = 1 + 6z^2 + 22z^4 + 68z^6 + 187z^8 + 470z^{10} + \ldots
\eeq
while for $P^r$ we have
\beq
A^r =
\prod_{m=1}^{\infty}(1 - z^{2m})^{-3}(1 + z^{2m})(1 + z^{4m})^2 = 1 + 4z^2 + 15z^4 + 44z^6 + 119z^8 + 292z^{10} + \ldots
\eeq
$k$ first terms of these expansions are exact in $P_k$ (checked up to $k=10$) and more patterns emerge with growing $k$.

\section{Conclusions}
\noindent
It is pointed out that, unexpectedly, polynomials in the expansion of the free energy in the thermodynamical limit in inverse powers of $(z^2+x^8)$ have to be divisible by growing powers of $(1-x^4)$ what provides a powerful check on the calculations. The patterns that are suggested by the results are not easy to understand and justify. Several combinatorial algorithms  are proposed that make the calculations of the expansion in $x^4$ and $z^{-2}$ feasible also for large powers, in the present case up to $x^{88}$ and $z^{-60}$.

\vspace{0.3cm} 
\noindent
 {\bf Acknowledgments:} 
 We thank Jacek Wojtkiewicz for discussions. We gratefully acknowledge Poland’s high-performance computing infrastructure PLGrid (HPC Centers: ACK Cyfronet AGH, PCSS) for providing computer facilities and support within computational grant no. PLG/2022/015410.Infrastructure.

\newpage

\centerline{\bf Appendix}

\vspace{2mm}

We quote here results for polynomials $M_k$ in a vector-like notation (for example $M_3 = -26x^8 + 20x^4 + 1$)
\bea
M_1 &=& [2]\nn\\
M_2 &=& [6]\nn\\
M_3 &=& [-26, + 20,+ 1]\nn\\
M_4 &=& [-127, + 71, + 8]\nn\\
M_5 &=& [672, - 898, + 204, + 46, + 2]\nn\\
M_6 &=& [3748, - 4166, + 588, + 236, + 22]\nn\\
M_7 &=&[ -21717,+ 40932, -20522, + 22, + 949, + 158, + 6]\nn\\
M_8 &=&[ -129520, + 214612, - 86971, - 6965, + 3771, + 963, + 77, + 1]\nn\\
M_9 &=& [790148, - 1918820, + 1483560, - 290292, - 75786, + 7222, + 4606, + 620, + 30]\nn\\
M_{10}&=&[4909146,-10797042,+7255634,-903474,-449562,-1444,+21452,+4194,+358,+8]\nn\\
M_{11}&=&[-30962104,+91955736,-95989312,+36734982,+648594,-2197024,-287998,+61648,\nn\\
&&+21974,+2930,+163,+2]\nn\\
M_{12}&=&[-197754011,+541893587,-509230276,+159559562,+16827439,-9743791,-2212687,+134119,\nn\\
&&+112320,+20670,+1840,+68]\nn\\
M_{13}&=&[1276651444,-4481462006,+5888629732,-3336912386,+507888490,+182860044,-23454082,\nn\\
&&-13707312,-907960,+373298,+113368,+14924,+1018,+22]\nn\\
M_{14}&=&[8318116584,-27285459280,+32957952130,-16392095652,+1420088296,+1094499026,-29149932,\nn\\
&&-71932170,-9874482,+1086016,+610844,+107522,+10576,+504,+6]\nn\\
M_{15}&=&[-54634033876,+221269623968,-350446626494,+262519589328,-79805854962,-4861680892,+5602322995,\nn\\
&&+694349132,-263792358,-74494034,-3086437,+2210278,+600766,+82994,+6436,+226,+1]\nn\\
M_{16}&=&[-361378718275,+1380363366137,-2037841061807,+1384516957207,-339726098694,-55777196860,\nn\\
&&+25372379516,+5805545546,-897872662,-441674754,-47589784,+7375250,+3339024,+597356\nn\\
&&+63150,+3786,+88]\nn\\
M_{17}&=&[2405313770932,-11038725876010,+20459076402020,-19040080044628,+8543166176964,\nn\\
&&-903999792052,-525107576520,+61065093712,+40791836918,+845399686,-1929749934,\nn\\
&&-415522266,-9391836,+12526844,+3342076,+479050,+42310,+1998,+30]\nn\\
M_{18}&=&[16098926860110,-70172478945796,+122441621521438,-105355508621080,+41389850278064,\nn\\
&&-1648391207328,-3057071838208,+68416975072,+223799676324,+22196957534,-8094930430,\nn\\
&&-2677483506,-236138658,+44220938,+18826588,+3425970,+392744,+27836,+994,+8]\nn\\
M_{19}&=&[-108290356982558,+555344393668748,-1178908721924776,+1310506856226150,-773054140102735,\nn\\
&&+187057690111770,+23737054642583,-15340462008456,-2164035828619,+906559319200,+230416622979,\nn\\
&&-9994439316,-13061798808,-2320605406,-36934929,+71656310,+19005030,+2868946,\nn\\
&&+281288,+16690,+446,+2]\nn\\
M_{20}&=&[-731709152575266,+3583691663824842,-7215834429918597,+7511351606333315,-4020817476109895,\nn\\
&&+754624227467905,+199817398773030,-68254814924066,-17763048105609,+3420212152505,\nn\\
&&+1483524029646,+64084853768,-61312916779,-15954041785,-1297876625,+258289635,\nn\\
&&+107459861,+20251417,+2492448,+203458,+9669,+187]\nn\\
M_{21}&=&[4964332892833076,-28133197238821276,+67300104285722984,-86967841632955304,+63409242872814544,\nn\\
&&-23227587332928562,+1282793709832898,+1633811587499322,-141181775899838,-128951641183966,\nn\\
&&-20889333998,+7497910244412,+1233681714920,-139104499720,-83484077492,-13561032290,\nn\\
&&-189970950,+406807266,+110581872,+17543456,+1891370,+133302,+5220,+68]\nn
\eea
\bea
M_{22}&=&[33806425311358576,-183784999141304612,+419441020752357056,-512186890035532240,\nn\\
&&+345770183378931912,-109630452562120848,-1850647584517146,+9156720168793754,\nn\\
&&+6423972242588,-712904992664848,-58479322658924,+35025628260858,+9163417307794,\nn\\
&&-19815098902,-420731560818,-97058895924,-7619414314,+1451363340,+622280014,\nn\\
&&+121910606,+16074882,+1472434,+87116,+2672,+22]\nn\\
M_{23}&=&[-231002135440625862,+1433527032947567734,-3815298343135201792,+5618371348540121274,\nn\\
&&-4871765425756103964,+2362188884960645290,-442366800051964037,-101294787760528598,\nn\\
&&+43417616564427070,+7812114469414480,-2976770019731457,-749191359290512,+77875540897627,\nn\\
&&+54218018148624,+6153037058657,-1185771757596,-530935259095,-81477374178,-1480726011,\nn\\
&&+2309686974,+651875220,+109083958,+12763607,+1037766,+52682,+1280,+6]\nn\\
M_{24}&=&[-1583399340278954651,+9460737607397274677,-24134353285045153077,+33813801413498875831,\nn\\
&&-27501553341656639662,+12050255919780550394,-1607301963657594836,-726904753275546804,\nn\\
&&+184680190402757457,+60174098606851379,-11473990972644543,-5009669037017009,+12735920920357,\nn\\
&&+285895458704795,+53486341877765,-2051140106911,-2803468862361,-598420478189,-48109704122,\nn\\
&&+7935239290,+3627508821,+744530445,+104559030,+10596356,+740911,+31323,+573,+1]\nn\\
M_{25}&=&[10884693554643712676,-73408493920695886416,+215138537745134201932,-355570454909753125570,\nn\\
&&+357103032988949049906,-214328467366164680836,+64567557020250901584,+228548135860027346,\nn\\
&&-5265821888538578210,+253687998047700046,+427429310958499346,-501271321184990,\nn\\
&&-27408275155779494,-3659877703088848,+943054815105740,+359954610770776,+30565891166372,\nn\\
&&-8899098709344,-3360517710786,-506170836250,-13059082862,+13002527194,+3882970500,\nn\\
&&+684708784,+86427172,+7894994,+489364,+17482,+238]\nn\\
M_{26} &=& [75023767231490887760,-488660853060009343032,+1377925415966711605784,-2178315228912642170888,\nn\\
&&+2070673377503681101816,-1148980256371538905758,+292961540923992725710,+26341406176661537010,\nn\\
&&-28304903123365607242,-1030596163346967024,+2333973697794365800,+191703897534125510,\nn\\
&&-133844798062631704,-30442763983477720,+2606201078299228,+2071780266912860,+311510001837782,\nn\\
&&-22229279045734,-18311342994990,-3760877228382,-318456407120,+41528232630,+21240981966,\nn\\
&&+4584117448,+684494924,+75675474,+6068030,+323702,+9470,+88]\nn\\
M_{27}&=&[-518387340284430020982,+3775214669775049458940,-12080269916586873168565,\nn\\
&&+22140347000608754568532,-25260985204632789248215,+18023658430925469281484,\nn\\
&&-7329903378160861110362,+986223728074131506460,+405624770544599070446,-123088111078947203624,\nn\\
&&-30082516832441395322,+9605253074632342476,+2605393325962197790,-353394764167166122,\nn\\
&&-206701805068231974,-13272731638989550,+8193340046502774,+2333066803917066,+150374885081637,\nn\\
&&-62190085554868,-21382632293001,-3221034789466,-116895226723,+72410084592,+23228937643,\nn\\
&&+4328229458,+585088983,+59075406,+4283674,+201146,+4830,+30]\nn\\
M_{28}&=&[-3590108432622951335298,+25316711368627966131280,-78193774533394538577742,\nn\\
&&+137665082605576918870268,-149668962863483902296618,+100129697659420156389018,\nn\\
&&-36473851175828928140444,+2827989074975941570096,+2635617061417455768228,-484642268458852715614,\nn\\
&&-215683514441037098382,+36180049358445857954,+17516655316608837166,-499261046232278414,\nn\\
&&-1157452959278185582,-161390856308545356,+30521581731227784,+14347371277456244,+1820693303869872,\nn\\
&&-181349647442526,-118851006417810,-23981903472220,-2184932637755,+204318017033,+124188237123,\nn\\
&&+28381423145,+4492794909,+536972689,+48246770,+3082978,+123720,+2364,+8]\nn\\
M_{29}&=&[24916720841121272335455,-194872736057505425718283,+676021920366590339436897,\nn\\
&&-1360614197832345026365699,+1737752621990791202659455,-1434497151968629767087049,\nn\\
&&+727833639756935607952019,-178359188031595289400347,-13902310991614079140589,\nn\\
&&+17152335076288853934005,-62341671204066640193,-1457875370765181192375,-16446304149487078165,\nn\\
&&+99084515754960031017,+11710272000335370753,-4468603849628970737,-1374367733467125552,\nn\\
&&-20272394137876164,+63544611331515132,+14930319863373106,+761005460460104,-419902307917166,\nn\\
&&-136622370428724,-20941545546944,-1005546596232,+394440900824,+139191902302,+27445654000,\nn\\
&&+3956479826,+435716300,+35938242,+2071258,+72244,+1084,+2]\nn
\eea

\end{document}